\documentclass[12pt]{article}
\usepackage{amssymb, amsfonts, graphicx, hyperref, psfrag, epsfig}
\newcommand{\be}{\begin{equation}}
\newcommand{\ee}{\end{equation}}
\newcommand{\bea}{\begin{eqnarray}}
\newcommand{\eea}{\end{eqnarray}}
\newcommand{\ba}{\begin{array}}
\newcommand{\ea}{\end{array}}

\csname @addtoreset\endcsname{equation}{section}

\begin{document}

\begin{titlepage}

\title{\bf\Large NGN,  QCD$_2$  and  Chiral Phase Transition  \\from  String Theory\vspace{18pt}}
\author{\normalsize
 Yi-hong Gao,  Wei-shui Xu and  Ding-fang Zeng \vspace{12pt}\\
{\it\small Institute of Theoretical Physics}\\ {\it\small P.O. Box 2735, ~Beijing 100080, ~China}\\
{\small e-mail: { \it gaoyh@itp.ac.cn,~\it wsxu@itp.ac.cn,~\it
dfzeng@itp.ac.cn}} }

\date{}
\maketitle \voffset -.2in
\vskip 1cm \centerline{\bf Abstract}
\vskip .4cm

We construct a D2-D8-$\bar{D8}$ configuration in string theory, it
can be described at low energy by two dimensional field theory. In
the weak coupling region, the low energy theory is a nonlocal
generalization of Gross-Neveu(GN) model which dynamically breaks the
chiral flavor symmetry $U(N_f)_L \times U(N_f)_R$ at large $N_c$ and
finite $N_f$. However, in the strong coupling region, we can use the
SUGRA/Born-Infeld approximation to describe the low energy dynamics
of the system. Also, we analyze the low energy dynamics about the
configuration of wrapping the one direction of D2 brane on a circle
with anti-periodic boundary condition of fermions. The fermions and
scalars on D2 branes get mass and decouple from the low energy
theory. The IR dynamics is described by the $QCD_2$ at weak
coupling. In the opposite region, the dynamics has a holographic
dual description. And we have discussed the phase transition of
chiral symmetry breaking at finite temperature. Finally, after
performing T-duality, this configuration is related to some other
brane configurations.

\vskip 2.cm \noindent May ~2006
\thispagestyle{empty}
\end{titlepage}

\newpage
\section{Introduction}

The ads/cft correspondence is one of the realizations of the
holographic principle. It relates string theory on some background
to $N=4$ SYM theory on the boundary \cite{Jm,Db}, and for a review
see \cite{Oa}. Inspired by this correspondence, one naturally asks
how to construct a holographic dual  model of the given
four-dimensional field theory and to study some interesting low
energy physics such as confinement and spontaneous chiral symmetry
breaking. Following the adf/cft, some new methods \cite{Ew,Po,Kl,Ma}
for studying strongly coupled gauge theories have been developed. In
\cite{Ew}, Witten proposed a construction of the holographic dual of
four-dimensional pure $SU(N_c)$ YM theory. In his construction, the
$N_c$ D4-branes wrap on a circle with antiperiodic boundary
conditions for the fermions living on the D4-branes. The
antiperiodic boundary conditions give a mass to fermions and scalars
on the D4-branes at tree level and one loop, respectively. Below
this mass scale the dynamics on the D4-branes can be described by
the pure $U(N_c)$ YM theory. But in such construction, there are no
flavor degrees of freedom. Observing this, Karch and Katz introduced
flavor branes to give  a construction of holographic models with
flavors \cite{Ka}. The SUGRA/Born-Infeld approximation can be used
to describe aspects of large $N_c$ gauge theories with finite $N_f$
flavors from the holographic point of view. There are many
developments along this line \cite{Ts1,Jb,Mk,Ts,Ts2,Cm}. For
example, in \cite{Mk}, M.Kruczenski, D.Mateos, R.C.Myers and
D.J.Winters(KMMW) considered the D4-D6-brane configuration, and in
\cite{Ts}, T.Sakai and S. Sugimoto(SS) proposed the
D4-D8-$\bar{D8}$-branes model and wished to obtain a better
understanding of $QCD_4$ with flavors from holographic point of
view. In the SS model, the global $U(N_f)_L \times U(N_f)_R$ chiral
flavor symmetry in $QCD_4$ is induced by the gauge symmetry of the
$N_f$ D8-$\bar{D8}$-brane pairs. SS applied SUGRA approximation to
analyze chiral symmetry breaking and compute the meson spectrum at
zero temperature.  In recently papers \cite{Js1,Ap,Dm,Pb,Tal}, the
authors studied the phase transition in a finite temperature using
the dual description. They found that at sufficiently low
temperature the chiral symmetry is broken, while for temperature
large than the critical value, the symmetry is restored. And this is
a first order phase transition between broken and unbroken chiral
symmetry phase. Also in \cite{Karch}, the chiral phase transition is
studied in strongly-coupled large $N_c$ N = 4 SYM theory on a
3-sphere coupled to a finite number $N_f$ of massive $N = 2$
hypermultiplets in the fundamental representation of the gauge
group.

Recently, in \cite{Ea}, Antonyan et al. studied the D4-D8-$\bar{D8}$
brane configuration where the D4-branes is not compactified on a
circle. In the weak 't Hooft coupling region, the low energy
dynamics of this D-brane configuration in string theory is described
by the nonlocal version of Nambu-Jona-Lasinio(NJL) model \cite{NJL}.
The NJL model is exactly solvable at large $N_c$  and dynamically
breaks the chiral flavor symmetry at weak 't Hooft coupling and is
non-renormalizable due to the four Fermi interaction. In the strong
coupling parameter region, the SUGRA approximation is reliable. By
varying the parameters, the D-brane configuration can interpolate
between NJL and $QCD_4$.

Along this line, we are mainly interested in the two dimensional
physics for example the chiral symmetry breaking in this paper. We
know that the two-dimensional version of the NJL model is the
Gross-Neveu(GN) model \cite{Gr}, which is superrenormalizable and
asymptotical free without gauge fields. Since it is a vector model
\cite{Mo}, it can be exactly solvable at large $N_c$ and finite
$N_f$. Also this model breaks the chiral symmetry dynamically and
generates a mass for quark via dimensional transmutation
\cite{Co}. In this paper we study a D2-D8-$\bar{D8}$-brane
configuration in IIA string theory. In this brane configuration,
the two dimensional theory has $U(N_f)_L \times U(N_f)_R$ chiral
symmetry which is induced from the gauge symmetry. At weak
coupling, the low energy dynamics of $q_L, q_R$ is described by
the non-local GN(NGN) model. We can find that the fermions get
mass and  the chiral symmetry $U(N_f)_L \times U(N_f)_R$
dynamically breaks to the diagonal subgroup $U(N_f)$. In strong
coupling region, however, we can use the SUGRA to describe the
dynamics and can get the similar results. Though the GN is a
two-dimensional toy model, it shares many same interesting
properties with the $QCD_4$. Hence its string realization can give
us better understanding of the four-dimensional physics,
especially some nonperturbative aspects, and also give us some
hints about how to embed the 4D physics into sting theory and
obtain some holograhic description. We will also consider another
case that's to compactify one direction of D2-branes with
antiperiodic boundary conditions for fermions on the D2-branes. In
this case, it will introduce another scale the radius of a circle
except for the distance between the parallel D8-$\bar{D8}$-brane
pairs in the low-energy theory. Thus, we need to analyze the
parameters space again. At weak coupling, the low energy dynamics
can be described by the two-dimensional $QCD_2$. However, in the
opposite region, the low energy dynamics has a holographic
description of $QCD_2$ by the supergravity. Also, we study the
chiral phase transition at high temperature. There is a critical
temperature, under it, the chiral symmetry is broken. But beyond
this temperature, the chiral symmetry is restored. For this brane
configuration, if performing T-duality along the compactified
direction, we can get the D1-D9-$\bar{D9}$-brane configuration
\cite{Su}. The low energy dynamics on D1-brane with $N_c=1$ is a
two-dimensional QED theory(Schwinger model \cite{Sc}). The
two-dimensional QED is a exactly solvable and can apply some power
techniques for example bosonization to study the D-brane dynamics.
If doing a T-duality along the D1-brane worldvolume again, we get
the D0-D8-$\bar{D8}$-branes configuration and have a matrix
description of the dynamics of the D0-branes. There are other
D-branes configurations such as D2-D6-$\bar{D6}$ and
D2-D4-$\bar{D4}$-branes system \cite{Maria}. The intersectional
part is also two dimensions as the D2-D8-$\bar{D8}$ brane
configuration. In this paper, we mainly focus on the
D2-D8-$\bar{D8}$ brane configuration. But for the other brane
configuration such as D2-D6-$\bar{D6}$ and D2-D4-$\bar{D4}$, one
can do the same analysis as the D2-D8-$\bar{D8}$ brane
configuration in the following sections of this paper.

The organization of this paper is as follows. In section 2, we
explain the brane configuration and the open string spectrum of
the D2-D8-$\bar{D8}$-branes system and discuss the valid region of
the weak and strong-coupling approximation. And we study the
low-energy dynamics at weak coupling and can find it is described
by a non-local GN model which exhibits chiral symmetry breaking at
weak coupling at large $N_c$. At strong coupling we use
SUGRA/Born-Infeld approximation to study the low-energy dynamics,
the fermions get mass and chiral symmetry is broken. In section 3,
we give a generalization to the above configuration and compactify
the one of D2-brane coordinates on a circle with antiperiodic
boundary condition for the fermions on D2-branes, then the
fermions and scalars on D2-branes get mass and is decoupled from
the low energy theory. Also, we analyze the properties of this
model at weak coupling and strong coupling region and the chiral
phase transition in finite temperature. After performing
T-duality, this configuration can relate to some other brane
system such as the D1-D9-$\bar{D9}$, D0-D8-$\bar{D8}$-branes
system. Finally, we give some comments and discussions about the
results of this paper in section 4.

\section{D2-D8-{$\bar{D8}$} brane configuration}\label{config}

Now let us introduce  a  brane configuration in IIA string theory
that are $N_c$ D2, $N_f$ D8 and $N_f$ $\bar{D8}$-branes. The
extending directions of these branes are indicated in the following

\be
\begin{array}{ccccccccccccc}
 &0 &1 &2 &3 &4 &5 &6 &7 &8 &9\\
 D2: &{\rm x} &{\rm x} &{\rm x} &{} &{} &{} &{} &{} &{} &{}\\
 {D8,\bar{D8}}: &{\rm x} &{\rm x} &{} &{\rm x} &{\rm x} &{\rm x} &{\rm x} &{\rm x} &{\rm x} &{\rm
  x}
\end{array}
\label{con} \ee In this brane configuration, the $N_f$ D8 and
$\bar{D8}$-branes are parallel and separated by a distance $L$ in
the $x^2$ direction. The $N_c$ D2-branes intersect with the $N_f$
D8-$\bar{D8}$-branes along directions $(x^0,x^1 )$. All the others
$(x^3,x^4, ... , x^9)$ are the transverse directions to the
intersection part of this brane configuration. Here the description
of the brane configuration is classically , as we will see in the
following sections the quantum fluctuation can change this classical
picture.  In this D-branes system, the particle spectrum can be
created by the open 2-2, 8-8, $\bar{8}$-$\bar{8}$, 8-$\bar{8}$, 2-8
and 2-$\bar{8}$ strings which ends on the respective D-branes. The
massless spectrum generated by the p-p open strings(where p denotes
2, 8, $\bar{8}$) have worldvolume gauge fields $A_\mu$($\mu$=
0,...,p), scalar fields $\phi^i$ (i=p+1,...,9) which describe the
transverse fluctuation of D-branes. Due to supersymmetry on D-brane,
there exists corresponding fermions on the D-brane worldvolume. The
fields of the gauge theories on the worldvolume can be obtained by
dimensional reduction of $N=1 U(n)$ ten dimensional SYM. All these
fields transform under the adjoint representation of the gauge group
$U(n)$. In this D-brane configuration, the $n$ is equal to $N_c$ for
the 2-2 open string. But for the 8-8 and $\bar{8}$-$\bar{8}$ case,
$n$ is chosen to $N_f$. To the 8-$\bar{8}$ open string, it may
create tachyon mode. The mass of this ground state is given by
$m^2=(\Delta x^2/2\pi\alpha')^2-1/2\alpha'$, where $\Delta x^2$ is
the distance between the D8-branes and $\bar{D8}$-branes. If the
distance $\Delta x^2>\sqrt{2}\pi l_s$, then one may think that this
mode is massive and decouple from the low energy physics. We will
give some discussions about the fundamental string in the background
(\ref{metric}) in the subsection \ref{ssss}.

As for the 2-8 and 2-$\bar{8}$ open strings, in the NS-sector the
massless spectrum depends on the $\#ND$ numbers of corresponding
D-branes of open string ending on. From (\ref{con}) we can get
$\#ND=8$, the vacuum zero energy is $1/2$. Thus, we can see that all
the states in NS-sector are massive, in the low energy theory, the
NS-sector is decoupled. For the R sector, since the vacuum zero
energy always vanishes, the ground state is massless. Then, the
ground states in the R-sector consist of positive and negative
chirality spinor representations of the Lorentz group $SO(1,1)$. Due
to the GSO projection, one of these two spinors is projected out. We
can choose the positive chirality spinors as physical states created
by the 2-8 strings and denote the fields as $q_L$ localized at
$x^2=-L/2$, which transform in the fundamental representation
($N_c$, $N_f$, 1) of the symmetry group $U(N_c)\times U(N_f)_L\times
U(N_f)_R$. The left-hand weyl fermion $q_L$ is complex since the 2-8
open strings have two orientations. Similarly, the massless
right-hand weyl fermions $q_R$ can be created by the 2-$\bar{8}$
open strings and locate at $x^2=L/2$, which transform in the
fundamental representation ($N_c$, 1, $N_f$) of the symmetry
$U(N_c)\times U(N_f)_L\times U(N_f)_R$.

We need to mention that the component $A_2$ of the gauge field
$A_M(M=0,1,2)$ and the scalar fields $\phi^i$ on the D2-branes
worldvolume can couple the quarks $q_L$ and $q_R$. But because of
the shift symmetry $ \phi^i\rightarrow \phi^i + const $ and gauge
invariance, they can only couple to the other massless modes through
irrelevant operators. In the low energy theory, the irrelevant
coupling don't play an important role. The fermions on D2-branes
also don't couple with quarks. Thus, the low energy dynamics on the
two-dimensional intersection is decoupled with the IR dynamics of
the scalars $\phi^i$, $A_2$ and 2-2 strings fermions on the
D2-branes worldvolume.

In the following, we are interested in the physics of intersectional
part of D-branes. Hence, we give a summary of all the massless modes
and their transformation under gauge symmetry in the $1+1$
intersectional dimensions. They can be listed in the following table
\ref{sdfdfs}.
\begin{table}[h] \tabcolsep0.33in
 \centering{
\begin{tabular}{|c|c|c|c|}
 \hline field &SO(1,1) &SO(8) & $U(N_c)\times
U(N_f)_L\times
U(N_f)_R$ \\
\hline   $A_\mu$   &2      &1       &(adj, 1, 1 )\\
\hline
  $q_L$ &$1_+$ &1 & $(N_c, N_f, 1)$\\
  $q_R$ &$1_-$ &1 & $(N_c, 1, N_f)$\\
  \hline

\end{tabular}
}\caption{the massless degrees of freedom on the intersection of
this branes configuration} \label{sdfdfs}
\end{table}

In the following studying, we are interested in this system at the
energy much below the string scale and in the limit $g_s\rightarrow
0$, $N_c\rightarrow \infty$ with $g_sN_c$ and $N_f$ held fixed. In
this brane configuration, the $D_2$-branes worldvolume theory is a
three dimensional $U(N_c)$ gauge theory with the gauge coupling
constant $g_3^2 = g_s/l_s$ \cite{Po1}. The 't Hooft coupling
constant can be defined as $\lambda = g_3^2 N_c/4\pi=g_sN_c/4\pi
l_s$. Since the coupling $g_3^2$ has units of mass, then the thing
is same for 't Hooft coupling constant. Now we can analyze the
coupling constant parameter space. As in \cite{It}, we can introduce
energy scale $U\equiv r/\alpha'$, at this energy scale, the
effective dimensionless coupling constant in three dimensional gauge
theory is $g^2_{eff}\approx\lambda/U$. In this brane configuration,
the energy scale is the distance $L$ between D8 and $\bar{D8}$
brane. Thus, in the region

\be l_s \ll L\ll \frac{1}{\lambda} \label{weak} \ee the physics that
we are interested lies at a distance scale much larger than the
string scale $l_s$, we can neglect the string effects. The effective
coupling constant $g^2_{eff}\approx\lambda L\ll 1$, the coupling is
weak and the perturbative calculation can be trusted in this region.
We can use the perturbative theory to describe the dynamics of
quarks $q_L$ and $q_R$. If $L\rightarrow \infty$, the
$g^2_{eff}\rightarrow \infty$, which means the interaction between
the left-hand quarks and the right-hand quarks becomes strong with
increasing the distance $L$. However, if let $L\rightarrow 0$, the
effective coupling constant $g^2_{eff}\rightarrow 0$. When the 't
Hooft coupling increases into the region
\bea l_s\ll \frac{1}{\lambda} \ll L  \nonumber \\
\frac{1}{\lambda} \ll l_s \ll L  \label{strong} \eea the effective
coupling constant becomes large and the interaction between $q_L$
and $q_R$ becomes strong. We can't use the above perturbative way to
deal with it, instead in this region we can use the
SUGRA/Born-Infeld approximation to study the low energy dynamics of
the brane system. In the subsection \ref{ssss}, the IR physics of
the brane system in the strong coupling region can be investigated
by analyzing the dynamics of D8-branes in the near-horizon geometry
of the D2-branes.

\subsection{Weak coupling region}\label{wcr}

In this subsection, we discuss the D2-D8-$\bar{D8}$-brane
configuration in the weakly coupled region (\ref{weak}). If the
distance $L$ between D8 and $\bar{D8}$-branes is approximate to
zero, there are no attractive force interaction between D8 and
$\bar{D8}$-branes. In this case, we only need to consider a single
intersectional branch between $N_c$ D2-branes and $N_f$ D8-branes.
Then the effective action of the quark $q_L$ and U($N_c$) gauge
field $A_M$ is given by\footnote{In this paper, we choose the two
dimensional $\gamma^i$ matrices as $\gamma^0 = i\sigma^2$,
$\gamma^1=\sigma^1$ and $\gamma^5=\gamma^0\gamma^1=\sigma^3$ where
the $\sigma^i$ are the Pauli matrices.}

\be S =
\int{d^3x\left[-\frac{1}{4g^2_3}F^2_{MN}+i\delta(x^2)q^\dag_L(-D_0+D_1)q_L\right]}
\label{effective}\ee where the indices $M=0,1,2$, the covariant
derivative $D_0$ and $D_1$ means $D_\mu=\partial_\mu-iA_\mu$
choosing $\mu=0, 1$, respectively. And the branes intersection is
located at $x^2= 0$. After integrating the gauge field $A_M$ and
approximating to first order about the gauge coupling $g^2_3$ in the
gauge field Lagrangian and in the Feynman gauge, the effective
action of $q_L$ is \be S_{eff} = i\int d^2x
 q^\dag_L(-\partial_0+\partial_1)q_L  \label{effective2}\ee Thus, we can see
, if up to first order correction, the effective action of left-hand quark is free. But if
concluding the high order correction, the quarks $q_L$ will
interact each other.

In the following we will discuss the two branches intersection case of the D2-D8-$\bar{D8}$-branes.
The distance between D8 and $\bar{D8}$-branes is $L$, and they locate
at $x^2=-L/2$ and $x^2=L/2$ respectively. Then the low energy
effective action on the intersection of branes \be
\begin{array}{lll} S =
&\int d^3x \left(-\frac{1}{4g^2_3}F^2_{MN}\right) \\
&+\int
d^2x\left(i\delta(x^2+L/2)q^\dag_L(-D_0+D_1)q_L+i\delta(x^2-L/2)q^\dag_R(-D_0-D_1)q_R\right)
\end{array}
\label{effective3}\ee Integrating out the gauge field in the single
gluon exchange approximation we get \be
\begin{array}{lll} S_{eff}=&
i\int d^2x\left[q^\dag_L(-\partial_0+\partial_1)q_L +
q^\dag_R(-\partial_0-\partial_1)q_R\right] \\  & ~~~~~~~~~ +
\frac{g^2_3}{4\pi}\int d^2xd^2yG(x-y,L)\left[q^\dag_L(x)\cdot
q_R(y)\right]\left[q^\dag_R(y)\cdot q_L(x)\right]\end{array}
\label{effective4} \ee where the 2+1 dimensional propagator is
$G(x,L)=\frac{1}{\sqrt{L^2+x_\mu x^\mu}}$. The color indices in
(\ref{effective4}) are contracted in each term in brackets
separately, while the flavor ones are contracted between $q_L$ from
one term and $q^\dag_L$ from the other. Thus the quantities in
square brackets are color singlets and transform in the adjoint
representation of $U(N_f)_L$. This is nothing but the nonlocal
generalization of the GN model.

Now we can study the nonlocal Gross-Neveu model(NGN) at large
$N_c$, and introduce a complex scalar field $T(x,y)$ which
transforms as fundamental representation $(1,\bar{N_f}, N_f)$
under the symmetry group $U(N_c)\times U(N_f)_L\times U(N_f)_R$.
Then the effective action (\ref{effective4}) becomes as follows
\bea
  \begin{array}{lll} S_{eff}=& i\int d^2x
\left[ q^\dag_L(-\partial_0+\partial_1)q_L + q^\dag_R(-\partial_0-
\partial_1)q_R \right] \\ & + \int d^2xd^2y
\left[-\frac{N_c}{\lambda}\frac{T(x,y)\bar{T}(y,x)}{G(x-y,L)}+\bar{T}(y,x)q^\dag_L(x)\cdot
q_R(y)+T(x,y)q^\dag_R(y)\cdot q_L(x)\right]\end{array}
\label{effective5}\eea where the scalar field is \be
T(x,y)=\frac{\lambda}{N_c}G(x-y,L)q^\dag_L(x)\cdot q_R(y)\label{t}
\ee In order to investigate the chiral symmetry breaking,
we need to calculate the vacuum expectation of the scalar field
$T(x,y)$. As done in \cite{Ea}, we will integrate out the fermion in
the large $N_c$ limit. The result effective action can get
correction only from the one loop diagrams with an arbitrary number
of external $T(x,y)$, $\bar{T}(x,y)$ fields. Thus, after dropping
some overall factor, the effective potential in the Euclidean
spacetime is \be V_{eff}=\int
d^2xT(x)\bar{T}(x)\frac{\sqrt{x^2+L^2}}{\lambda}-\int\frac{d^2k}{(2\pi)^2}
ln\left(1+\frac{T(k)\bar{T}(k)}{k^2}\right) \ee where $T(k)$ is the
Fourier transform of scalar field $T(x)$. The effective potential is
very similar to the one in \cite{Ea} and the only difference is the
integrating dimension.

From the effective potential, we can derive the equation of motion
that is \be \int d^2x\frac{\sqrt{x^2+L^2}
}{\lambda}T(x)e^{-ikx}=\frac{T(k)}{k^2+T(k)\bar{T}(k)}
\label{vvv}\ee There is a trivial solution of this equation,
$T(x)=0$. This means that the chiral symmetry is unbroken under this
vacuum. In following, we want to see if there exists other
nontrivial solutions of this equation which break the chiral
symmetry. As the same in \cite{Ea}, we can separate two regions. The
first is the linear region, $ T(k)\bar{T}(k)\ll k^2 $. In this
region, the equation becomes as the following form \be
\bigtriangledown^2\left[\frac{\sqrt{x^2+L^2}}{\lambda}T(x)\right]+T(x)=0
\label{eq}\ee where the $\bigtriangledown^2$ is the two-dimensional
Euclidean Laplacian. Using the numerical analysis, we can find there
exists nontrivial solution $T(r)$. The asymptotic behaviors of the
$T(r)$ are as follows. If  the $|x|\rightarrow 0$, then $T(x)$ leads
to constant. Otherwise $|x|\rightarrow \infty$, the $T(x)\rightarrow
0$. Thus there exists non-zero $T(x)$ solution which means that the
chiral symmetry is broken.

In below, we follow the argument in \cite{Ea} to discuss this
differential equation. We define $r=\sqrt{x^2}$ and $
F(r)=\sqrt{r^2+L^2}~ T(r)=\frac{\lambda}{N_c}<q^\dag_L(x)\cdot
q_R(0)>$, then the equation (\ref{eq}) becomes \be
F''(r)+\frac{1}{r}F'(r)+\frac{\lambda F(r)}{\sqrt{r^2+L^2}}=0
\label{eq1}\ee  This equation is difficult to solve analytically.
For $r\ll L$, the equation in term of the variable $\rho\equiv
r\sqrt{\lambda/L}$ can be written as \be
F''(\rho)+\frac{1}{\rho}F'(\rho)+ F(\rho) =0 \ee then the solution
is the Bessel function $J_0(\rho)$. If $\rho$ approaches to zero,
$J_0(\rho)$ leads to constant which is valid in the region $r\ll L$,
corresponding to $ \rho \ll \sqrt{\lambda L}\ll 1$. As for $r\gg L$,
the equation (\ref{eq1}) with the variable $\sigma \equiv \lambda r$
reaches the following form \be
F''(\sigma)+\frac{1}{\sigma}F'(\sigma)+ \frac{1}{\sigma}F(\sigma)= 0
\ee The solution of equation at large $\sigma$ can be expressed as
$F(\sigma)\sim J_0(2\sqrt{\sigma})+Y_0(2\sqrt{\sigma})$, it
approaches to constant again at large $\sigma$. Thus, $F(r)$ is
constant both for $r\ll L$ and for $r\gg L$. One can think that
there are not any non-trivial variation for $r\sim L$, and the two
constants are the same to leading order in $\lambda L$. As in
\cite{Ea}, we argue that the solution of (\ref{eq1}) is simply
$F(r)=C$, where $C$ is a constant. Then scalar field $T(r)$ solution
is \be T(r)=\frac{C}{\sqrt{r^2+L^2}} \label{equ}\ee Its Fourier
transformation is $T(k) \simeq \frac{C}{k}$.

The constant $C$ is not determined by the linear analysis and must
match with the non-linear region \be T(k)\bar{T}(k)\gg k \ee In this
opposite region, the equation (\ref{eq}) reduces to \be
\frac{1}{\bar{T}(k)}=\int d^2x
\frac{\sqrt{x^2+L^2}}{\lambda}T(x)e^{-ikx} \label{equa} \ee It is
easy to see that there exists one solution that is\be
T(x)=\bar{T}(x)=A\delta^2(x) \label{eq2}\ee and the Fourier
transformation $ T(k) = \bar{T}(k)= A $, where the constant is
$A=\sqrt{\frac{\lambda}{L}}$. Then the constant $C$ can be
determined by matching the linear and non-linear region.  We can get
$\frac{C}{k^*}\simeq \sqrt{\frac{\lambda}{L}}$ at the momentum
$k^*$. At this matching point $k^*$, T(k) and $\bar{T}(k)$ satisfy
$T(k^*)\bar{T}(k^*)\simeq {k^*}^2$. Then we find that the constant
$C$ satisfies \be
\begin{array}{lll}
 &A\sim k^*\sim \sqrt{\frac{\lambda}{L}} \\ & ~~~
 C\sim Ak^*\sim \frac{\lambda}{L}
\end{array}
\label{sss} \ee  Therefore, in the linear region $k\gg
\sqrt{\frac{\lambda}{L}}$, the equation takes the form $(\ref{eq})$
and the solution is given by $(\ref{equ})$. The quark-antiquark
condensate $<q^\dag_L(x) \cdot q_R(y)> \simeq \frac{N_c}{L}$ is
independent of parameters $r$ and $\lambda$ and approaches to zero
when $L\rightarrow \infty$. In the opposite nonlinear region $k\ll
\sqrt{\frac{\lambda}{L}}$, the equation is $(\ref{equa})$, and its
solution is given by $(\ref{eq2})$. The chiral condensate decrease
rapidly to zero. It shows that there exists non-trivial solutions of
the equation (\ref{vvv}), and therefore the nonzero chiral
condensate expectation value breaks chiral symmetry $U(N_f)_L\times
U(N_f)_R\rightarrow U(N_f)_{diag}$ in this non-local GN model.

\subsection{Strong coupling region}\label{ssss}

In this subsection we will analyze the dynamics of the chiral
fermions $q_L$ and $q_R$ in the strong coupling region
(\ref{strong}). In this region the perturbative analysis breaks
down, instead we will use the SUGRA/Born-Infeld approximation to
study the dynamics of D8-branes in the near-horizon geometry of the
$N_c$ D2-branes \cite{It}.

As in \cite{It}, the near-horizon geometry of the $N_c$ D2-branes is
given by \be ds^2=\left(\frac{\alpha'U}{R}\right)^{5/2}
\left(\eta_{\mu\nu} dx^\mu dx^\nu +\left(dx^2\right)^2 \right)
+\left(\frac{R}{\alpha'U} \right)^{5/2} \left(
\left(\alpha'dU\right)^2 +\left(\alpha'U\right)^2d\Omega_6^2 \right)
\label{metric}\ee where $\Omega_6$ is the angular directions in
$(3456789)$ and $U= r/\alpha'$ with transverse radial coordinate
$r$. The D2-branes also has a nontrivial dilation background \be
e^\phi
=\left(\frac{R^{25}}{(6\pi^2N_c)^4\alpha'^{25}U^5}\right)^{1/4}\label{dilaton}=\left(\frac{6\pi^2g_3^{10}N_c}{\alpha'^{10}U^5}\right)^{1/4}\ee
and the parameter R is given by
$R^5=6\pi^2N_c\alpha'^3g_3^2=6\pi^2N_cl_s^6g_s$.

In the following we set $\alpha' \equiv 1$, the D8-brane extend in the
$R^{1,1}\times R\times S^6$ directions. Then we take a D8-brane to
probe the geometry (\ref{metric})(For $N_f$ coincident D8-branes,
the results is same since we don't turn on the gauge fields on
D-branes and some bulk fields). The track of D8-brane forms a curve
$U=U(x^2)$ in the $(U,x^2)$ plane, whose shape is determined by the
equations of motion that follows from the DBI action of the
D8-brane. In the background (\ref{metric}) and (\ref{dilaton}), the
induced metric on the D8-brane is \be
ds^2=\left(\frac{U}{R}\right)^{5/2}\eta_{\mu\nu} dx^\mu dx^\nu
+\frac{R^{5/2}}{U^{1/2}}d\Omega_6^2
+\left(\left(\frac{U}{R}\right)^{5/2}\left(\frac{\partial
x^2}{\partial U}\right)^2 +\left(\frac{R}{U}\right)^{5/2}\right)dU^2
\label{metric1}\ee
 the action is \be S_{D8}\sim\int dx^2 U^{7/2}\sqrt{{1+ \left(\frac{R}{U}\right )^5}U'^2} \label{act}\ee
where the $U'= dU/dx^2$. Since the integrand of (\ref{act}) has no
explicit $x^2$ dependence, the equation of motion is \be
\frac{d}{dx^2}\left(\frac{U^{7/2}}{\sqrt{1+\left({\frac{R}{U}}\right)^5
U'^2}}\right)=0 \ee Then we can get the first order differential
equation\be \frac{U^{7/2}}{\sqrt{1+\left({\frac{R}{U}}\right)^5
U'^2}}= U_0^{7/2}\label{eee} \ee The solution $U(x^2)$ of
(\ref{eee}) is a shaped curve in the (U, $x^2$) plane. Here, we
choose the following boundary conditions.  For the $U \rightarrow
\infty$, the $x^2=\pm\frac{L}{2}$, and $U$ is equal to $U_0$ at
$x^2=0$. And the curve $U(x^2)$ is symmetric under the reflection
$x^2\rightarrow -x^2$. Solving the above for $U'$, the solution
$x^2(U)$ can be obtained as the integral form  \be
x^2(U)=\int_{U_0}^U\frac{dU}{\left(\frac{U}{R}\right)^{5/2}\sqrt{\left(\frac{U^7}{U_0^7}-1\right)}}
\label{dd}\ee To integrating this integral, we consider the
approximation $U/U_0\gg1$,then approximately the up bound of the
integral leads to infinity. Therefore we get \be x^2(U)= \frac{
R^{5/2}}{7U_0^{3/2}}\left(B(\frac{5}{7},\frac{1}{2})-B(\left(\frac{U_0}{U}\right)^7,\frac{5}{7},\frac{1}{2})\right)
\label{aa}\ee From the equation (\ref{aa}), we know the asymptotic
value $ L/2 = x^2(\infty) =\frac{R^{5/2}}{7
U_0^{3/2}}B(\frac{5}{7},\frac{1}{2}) $. At small $z$, the incomplete
Beta function can be expanded in the following form \be
B(z,a,b)=z^a\left(\frac{1}{a}+\frac{1-b}{a+1}z+\cdots\right)\ee Then
the form of the curve at large U is approximate obeyed to the
equation
 \be U^5 = \frac{R^{5/2}U_0^{7/2}}{5\left(L/2-x^2(U)\right)}
\label{ee} \ee Since the symmetry $x^2\rightarrow -x^2$, the part of
the D-brane at $x^2< 0$ is determined by $U(x^2)= U(-x^2)$. The full
D8-brane shape can be determined in background (\ref{metric}) and is
shown in the following fig. \ref{gn2}

\begin{figure}[h]
\centering{\includegraphics[scale=0.9]{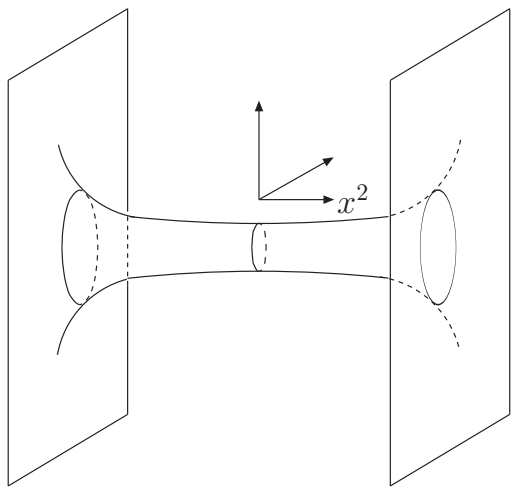}}
 \caption{ the shape of the D8-$\bar{D8}$-branes in the background (\ref{metric}).}
 \label{gn2}
 \end{figure}

We see that the configuration is deformed by the quantum effects.
In classical case, we have discussed in the above section that the
$N_f$ D8 and $N_f$ $\bar{D8}$-branes in this brane configuration
is sit at $x^2 =\frac{L}{2}$ and $x^2 = -\frac{L}{2}$
respectively. And they extend to the other directions include the
$U$ coordinate. They are classically not connected and the chiral
symmetry $U(N_f)_L\times U(N_f)_R$ is not broken. But here the
$N_f$ D8-branes and $N_f$ $\bar{D8}$-branes are joined into a
single D8-branes by a wormhole such as in \cite{Cg} and the chiral
symmetry $U(N_f)_R \times U(N_f)_L$ is dynamically broken to a
$U(N_f)_{diag}$.

We have seen there are two brane configuration, one is the separated
parallel brane configuration with fixed distance $L$, the other is
the connected one by the wormhole with the minimal $U$ value $U_0$.
We need to compare the energy density of these two configuration to
see which one is preferred. The energy density difference of the two
configurations is given by \be \begin{array}{lll}\bigtriangleup
E&\equiv E_{straight}-E_{curved} \\ &\sim \int_0^{U_0} (U - 0)dU +
\int_{U_0}^\infty \left(U-U(1-\frac{U_0^7}{U^7})^{-1/2}\right)dU
\approx 0.4 U_0^2 \end{array}\ee We find the energy density
difference $\bigtriangleup E$ is positive. This result means that
the connected configuration by the wormhole is preferred and the
chiral symmetry is likely to break.

If the above supergravity approximation is valid, the two conditions
must be satisfied. One is the dilation satisfies $e^\phi \ll 1$, and
the other is the curvature scalar that satisfied $\alpha'R\ll 1$.
Therefore the 't Hooft coupling constant must be satisfied \be U \ll
\lambda \label{bc}\ee which is extended by the eleven-dimensional
SUGRA. From (\ref{aa}), we have $U_0^3 \sim \lambda / L^2$. And in
the strong coupling region (\ref{strong}), $U_0$ obviously satisfies
the condition (\ref{bc}). Fixing $L$ and increasing $\lambda$ will
push $U_0$ further into the region of validity of supergravity. To
the opposite direction, decreasing $\lambda$ will make $U_0$ become
smaller, and when $L\simeq \frac{1}{\lambda}$ the curvature at $U_0$
becomes of order one and the supergravity description breaks down.
If continuing to decrease the $\lambda$ into the region $L \ll
\frac{1}{\lambda}$, the coupling becomes weak and the perturbative
description is reliable as in subsection \ref{wcr}.

Until now, we have analyzed the low energy dynamics of this brane
configuration in two different coupling region. In the weak coupling
region (\ref{weak}), the non-local GN description of the
two-dimensional dynamics is valid. And we find that the quarks get
mass (\ref{sss})and the chiral symmetry is dynamically broken.
However, in the stromg coupling region, SUGRA description is
reliable. We see that the energy density difference between  the
separated D8 and $\bar{D8}$-branes configuration and  the connected
D8-$\bar{D8}$-branes one by wormhole is positive. Thus the D8 and
$\bar{D8}$-branes is deformed to a connected curved D8-branes that
provides a geometric realization of chiral symmetry breaking. The
energy scale associated with chiral symmetry breaking in the
supergravity regime is $U_0\sim
\left(\frac{\lambda}{L^2}\right)^{1/3} $ which can be regarded as
the quark mass of $q_L$ and $q_R$ and correspond to $k^*\sim
\sqrt{\frac{\lambda}{L}}$ in (\ref{sss}) in weak coupling region.

From the above, we know that in the one region (\ref{weak}), the
field is valid, in the opposite region (\ref{strong}), the
supergravity is trusted. One may ask how to get some information
about the correspondence between these two sides. That means to find
the correspondence between the bulk fields and the boundary
operators. In below we give some superficial discussion following
from \cite{Ea}. We know that there are two kinds of bulk fields in
this configuration. One is the massless modes of the closed string
which exist even when the D8-branes are absent. The other is open
string fields on the D8-branes. Both two cases are coupled to quarks
$q_L$ and $q_R$. We consider first the open tachyon stretched
between the D8 and $\bar{D8}$-branes. This complex scalar field
which transforms as $(\bar{N_f}, N_f)$ under $U(N_f)_L\times
U(N_f)_R$ couples to the fermion via  Yukawa-type coupling \be
{\rm{L_1}}\simeq Tq^\dag_R\cdot q _L+\bar{T}q^\dag_L\cdot q_R \ee
Thus, the open string tachyon $T$, $\bar{T}$ is dual to the boundary
operators \be
\begin{array}{lll}
&T\leftrightarrow q^\dag_R\cdot q_L \\
&\bar{T}\leftrightarrow q^\dag_L\cdot q_R \end{array}\ee Therefore,
in addition to its curved shape (\ref{dd}), the connected curved
D8-brane describes a vacuum with non-zero expectation value
$<q^\dag_R\cdot q_L>$ will have a non-zero condensate of the
$8-\bar{8}$ tachyon $T$. Another bulk field is the scalar $(X^2)_L$
that parameterizes the location of the D8-branes in the direction
$x^2$, and  $(X^2)_R$ for the $\bar{D8}$-branes. Unlike the tachyon,
these scalars have the coupling \be {\rm{L_2}}\simeq
(X^2)_L\left(q^\dag_L\cdot q_Rq^\dag_R\cdot q_L
\right)_{(N_f^2,1)}+(X^2)_R\left(q^\dag_L\cdot q_Rq^\dag_R\cdot q_L
\right)_{(1,N_f^2)} \label{rr}\ee The quarks coupling $\rm{L_2}$
don't have interaction if the $L\rightarrow 0$. But for finite $L$,
there exists interaction between quarks (\ref{rr}). In the
near-horizon geometry, the coupling (\ref{rr}) implies that the
bulk-boundary correspondence is \be
\begin{array}{lll} &(X^2)_L\leftrightarrow\left(q^\dag_L\cdot
q_Rq^\dag_R\cdot
q_L\right)_{(N_f^2,1)}\\
&(X^2)_R\leftrightarrow\left(q^\dag_L\cdot q_Rq^\dag_R\cdot q_L
\right)_{(1,N_f^2)}\end{array} \label{rrr}\ee The chiral symmetry
breaking implies that the operators on the right-hand side of
(\ref{rrr}) have non-zero expectation value in our brane
configuration, and which leads that the scalars $(X^2)_L$, $(X^2)_R$
must have a non-zero expectation value.  This is the curved shape
$U(x^2)$ of the branes which is given asymptotically by (\ref{ee}).
And $x^2$ approaches to $L/2$ as $ U\rightarrow\infty$ like $L/2-x^2
\sim U^{-5}$. One can investigate the other bulk fields to find the
correspondent boundary operators. The details of the correspondence
is need to further study. In this paper, we don't expand to discuss
it in details and leave to study in the future.

\begin{figure}[h]
\centering{\includegraphics[scale=0.9]{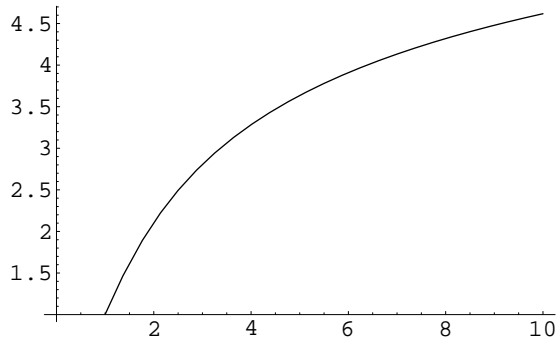}}
 \caption{the behavior of the length of curved string compared with the straight one in background (\ref{metric}).}
 \label{gn3}
 \end{figure}

We now turn to give some discussions to the $8-\bar{8}$ strings in
the background (\ref{metric}) as mentioned at the beginning of the
section \ref{config}. As we have analyzed in the above section if
the distance $L$ is smaller than the string length, the ground
string mode is a tachyon, otherwise it is massive. The
fundamental string is stretched between the D8 and $\bar{D8}$-branes
in the near-horizon geometry of the D2-branes (\ref{metric}) and is extending in
the directions $(x^0, x^2)$ \cite{Jr, Jm1,Br, Ea}. The induced metric
on the string worldvolume is given by \be ds^2 =
\left(\frac{U}{R}\right)^{5/2}
(dx^0)^2-\left({\left(\frac{U}{R}\right)^{5/2}+
\left(\frac{R}{U}\right)^{5/2}\left(\frac{dU}{dx^2}\right)^2}\right)
 (dx^2)^2 \label{induce} \ee Then we can compute the fundamental string Nambu-Goto
 action, and it is proportional to
 \be S_{NG}\sim -\int_{-L/2}^{L/2} dx^2
 \left(\frac{U}{R}\right)^{5/2}\left(1+\left(\frac{R}{U}\right)^5 U'^2\right)^{1/2}
 \ee
where the $U'\equiv \frac{dU}{dx^2}$. Since the integrand is not
explicit dependent on the $x^2$, then the equation of motion of
fundamental string is \be
\frac{d}{dx^2}\left(\frac{(U/R)^{5/2}}{\sqrt{1+\left(\frac{R}{U}\right)^5U'^2}}\right)=0\ee
If we choose the boundary conditions, $U'(x^2)=0$ and $U=U_0^F$ at
$x^2=0$, and $x^2= \pm\frac{L}{2}$ if the $U\rightarrow \infty$.
Then the first order equation is \be
\frac{(U/R)^{5/2}}{\sqrt{1+\left(\frac{R}{U}\right)^5U'^2}} =
(U_0^F/R)^{5/2} \label{jj}\ee We have noted that the $U_0^F$ is the
minimal value of U. From (\ref{jj}) and changing the integral
variable, the equation becomes the integral form \be
\begin{array}{lll} x^2(U)&=
\int_{U_0^F}^{U}\frac{dU}{(\frac{U}{R})^{5/2}\left(\left(\frac{U}{U_0^F}\right)^5-1\right)^{1/2}}
\\&=\frac{R^{5/2}}{(U_0^F)^{3/2}}\int_1^{\frac{U}{U_0^F}}dx
x^{-5/2}(x^5-1)^{-1/2}\end{array}\label{kk} \ee For the large U,
this means that U is larger than $U_0^F$, so the upper limit of the
integral (\ref{kk}) can be taken to infinity. Then the result of the
integral is \be x^2(U)=
\frac{R^{5/2}}{5(U_0^F)^{3/2}}\left(B(\frac{4}{5},\frac{1}{2})-B(\left(\frac{U_0^F}{U}\right)^5,\frac{4}{5},\frac{1}{2})\right)
\ee This means the distance $\frac{L}{2}=
\frac{R^{5/2}}{5(U_0^F)^{3/2}}B(\frac{4}{5},\frac{1}{2})$. For the
large $U$, the curve shape of the fundamental string in this
background is described by\be U^4\simeq
\frac{(U_0^F)^{5/2}R^{5/2}}{4(\frac{L}{2}-x^2(U))}\ee Also we can
compare the minima value of $U$ for the fundamental string and
D8-brane that is \be
\frac{U_0^F}{U_0}=\frac{7B(\frac{4}{5},\frac{1}{2})}{5B(\frac{5}{7},\frac{1}{2})}\approx
1.4 \label{wsxu}\ee Thus the minimal $U$ value of the fundamental
string is larger than the correspondent one of D8-brane and stays a
short distance with the minimal value $U_0$ of D8-branes. This
picture is same as in \cite{Ea}. We can compare the length of curved
fundamental string with the straight one. The length of curved is
proportional to \be L_{curved}\sim \int_1^x \frac{dx
 x^{5/4}}{\sqrt{x^5-1}} \label{ha1}\ee and the straight one is \be
L_{straight}\sim \int_1^x \frac{dx }{x^{5/2}\sqrt{x^5-1}}
\label{ha2}\ee Where the $x\equiv \frac{U}{U_0^F}$. And then the
behavior of the compared value $\frac{L_{curved}}{L_{straight}}$ is
shown in the figure \ref{gn3}. With increasing the value of $U$, the
compared value will become large. And the tachyon mode becomes
massive and decoupled. Under ours definition (\ref{ha1}) and
(\ref{ha2}), we find the behavior is different to the case of the
fundamental string in the near horizon geometry of $N_c$ D4-branes
in \cite{Ea}.

\section{A generalization of D-brane configuration}\label{gera}

In this section, just as did in \cite{Ew,Ts, Ea}, we generalize the
brane configuration of last section by wrapping the one direction
$x^2$ of D2-brane on a circle of radius $R_2$ with antiperiodic
boundary conditions for the fermion on the D2-branes worldvolume.
The antiperiodic boundary conditions break all the susy, and the
mass of the adjoint fermions and scalars are in general produced via
at tree level and one loop, respectively \cite{Js}. As discussed in
section \ref{config}, the trace parts of the scalar fields $A_2$ and
$\phi^i(i=3,\cdots,9)$ is still massless by shift symmetry. However,
since they can only couple to the other massless modes through
irrelevant operators, and they are not important in the low energy
theory. Below the mass of the fermions and scalars, the only
dynamical degree of freedom are $U(N_c)$ gauge fields and $N_f$
fermions in the fundamental representation of the gauge group
$U(N_c)$ as showed in section \ref{config}. We think the low energy
effective theory on two dimension is $QCD_2$ \cite{Gt}. But because
of an infinite tower of Kaluza-Klein modes of mass scale
$M_{KK}(M_{KK}\sim 1/R_2)$, at high energy the low energy dynamics
of this configuration can't be described by $QCD_2$ .

As in the previous section, we can analyze the coupling parameter
space. But in this case one more scale parameter, the radius $R_2$
of a circle for the $x^2$ direction, is introduced by the
compactified D2 branes. If the radius $R_2$ approaches infinite, it
can go back to the original brane configuration. We now will begin
to discuss the coupling parameter space. Firstly, in the region \be
l_s\ll L \ll R_2\ll \frac{1}{\lambda} \label{w} \ee the 't Hooft
coupling constant is small, and the low energy dynamics on the
D2-branes is weakly coupled. The compactified radius $R_2$ is finite
and the two-dimensional 't Hooft coupling constant
$\lambda_2=\frac{\lambda}{2\pi R_2}$ is also small, since the
effective coupling constant $ g_{eff} \sim L^2/(\frac{2\pi
R_2}{\lambda})\ll 1 $. Therefore, in this region the two-dimensional
theory is also weak coupled, and we can use the usual $QCD_2$
techniques to study this theory. Similar to the discussions
\cite{Ea}, for small $\lambda_2$, since the scale $\Lambda_{QCD_2}$
is much smaller than the dynamically generated quark mass $k^*\sim
\sqrt{\lambda/L}$, the low energy theory can separate into two
essentially decoupled parts. The one at distance scale $1/k^*$ is a
chiral symmetry breaking part that is described by the non-local GN
model as in section two. The second one is a confining part at the
distance $1/\Lambda_{QCD_2}$.

With increasing the 't Hooft coupling constant, the parameter
relation becomes \be l_s\ll L\ll \frac{1}{\lambda}\ll R_2
\label{ww}\ee We find the three dimensional effective coupling
constant $g_{eff} \sim L\lambda)\ll 1$, so the theory is weakly
coupled. And the two dimensional effective coupling constant is
still $L^2/(\frac{2\pi R_2}{\lambda})\ll 1$. As in region (\ref{w}),
the three-dimensional and two-dimensional coupling are both weak,
thus we can use the perturbative analysis.

Further increasing $\lambda$, the coupling parameter relation
reaches \be \frac{1}{\lambda}\ll l_s\ll L\ll R_2 \label{www} \ee The
't Hooft coupling $\lambda$ is large while the two dimensional
effective coupling constant $g_{eff}\sim L^2/(2\pi R_2/\lambda)$ can
be smaller or larger than one. If $L\lambda)\gg R_2/L$, then the
effective coupling, $g_{eff}\ll 1$, in this case the
three-dimensional dynamics must be described by SUGRA as in section
\ref{ssss}, however the two-dimensional one is still described by
$QCD_2$. In the opposite case, $g_{eff}\gg 1$, both the
three-dimensional and two-dimensional dynamics are strong coupling,
thus we must use the supergravity to study both the chiral symmetry
breaking and confinement.

In the following, we focus on the strong coupling region both the
two and three dimensional couplings. In this case, the supergravity
is trusted. As the same argument in \cite{Ew}, if compactifying the
one direction $x^2$ on a circle of radius $R_2$ with antiperiodic
boundary conditions for the fermions living on the D2-branes
worldvolume. we find that the near-horizon geometry metric of the
D2-branes is \bea &ds^2= \left(\frac{U}{R}\right)^{5/2}(-(dx^0)^2 +
(dx^1)^2 +
f(U)(dx^2)^2)+\left(\frac{R}{U}\right)^{5/2}(\frac{dU^2}{f(U)}+U^2d\Omega_6^2)
\label{met} \\ &e^\phi=
\left(\frac{6\pi^2N_cg_3^{10}}{U^5}\right)^{1/4}, ~~~f(U)=
1-\left(\frac{U_{KK}}{U}\right)^5 \label{me}\eea where we set the
$\alpha'\equiv1$,  the $R^5\equiv 6\pi^2N_cg_3^2$ and the constant
parameter $U_{KK}$. In order to avoid a singularity at $U=U_{KK}$,
the $x^2$ must be a periodic variable with \be x^2\sim x^2 + \delta
x^2, ~~~~ \delta x^2 = \frac{4\pi R^{5/2}}{5U_{KK}^{3/2}}
\label{antip}\ee we can define the Kaluza-Klein mass as \be
M_{KK}\sim \frac{1}{R_2}=\frac{2\pi}{\delta
x^2}=\frac{5U_{KK}^{3/2}}{2R^{5/2}} \ee Then the parameter constant
$U_{KK}$ satisfied \be U_{KK}^3= \frac{24\pi^2\lambda}{25R_2^2}
\label{uu} \ee The $U$ coordinate is restricted to $U\geq U_{KK}$ in
order to finite volume of $S^6$ and warp factor of $R^{1,1}$. And
for $U_0$ it is also larger than $U_{KK}$, which can see from the
equation (\ref{www}), (\ref{uu}) and $U_0\sim (\lambda/L^2)^{1/3}$.

We now will consider a probe D8-brane propagating in this metric
background. The D8-brane is extended in the directions
$R^{1,1}\times R\times S^6$, and we are interested in the solution
$x^2(U)$ in the coordinates plane $(x^2, U)$. The induced metric on
the D8-brane is \be \begin{array}{lll} ds^2 &=
\left(\frac{U}{R}\right)^{5/2}(-(dx^0)^2 + (dx^1)^2)\\
&+\left(\left(\frac{U}{R}\right)^{5/2}f(U)\left(\frac{dx^2}{dU}\right)^2+
\frac{1}{f(U)}\left(\frac{R}{U}\right)^{5/2}\right)dU^2+
\left(\frac{R}{U}\right)^{5/2}U^2d\Omega_6^2 \end{array}
\label{in2}\ee From the induced metric (\ref{in2}), we get the DBI
action of the D8-brane \be S_{D8} \sim \int dx^2
U\sqrt{\left(\frac{U}{R}\right)^5f(U)+ \frac{U'^2}{f(U)}}
\label{eqn}\ee where the $U'= \frac{dU}{dx^2}$. Since the integrand
of (\ref{eqn}) does not explicitly depend on $x^2$, we can obtain
the equation of motion as the energy conservation law \be
\frac{d}{dx^2}\left(\frac{U^6f(U)}{\sqrt{\left(\frac{U}{R}\right)^5f(U)+
\frac{U'^2}{f(U)}}}\right)= 0 \ee  Assuming the initial conditions
$U(0)=U_0$ and $U'(0)= 0 $ at $x^2=0$, the solution of this equation
of motion is obtained as \be x^2(U)=
U_0^{7/2}f(U_0)^{1/2}\int_{U_0}^U\frac{dU}{\left(\frac{U}{R}\right)^{5/2}f(U)\sqrt{U^7f(U)-U_0^7f(U_0)}}\label{aaa}
\ee For the approximation $U\gg U_0$, the integral (\ref{aaa}) can
integrate from $U_0$ to $\infty$, and $x^2(\infty)= L/2$. If
$U_0=U_{KK}$, then the
$x^2(\infty)\approx\frac{R^{5/2}}{5U_0^{3/2}}\sim \delta x^2$. In
this case, it means the $L\sim R_2$, and the D8-brane and the
$\bar{D8}$-brane are at antipodal points on the $S^1$ circle of the
$x^2$ direction. If $U_0\rightarrow \infty$, then the D8 and
$\bar{D8}$-branes approaches to infinity and disappears. The
distance $\frac{L}{2}= x^2(\infty)$ is a monotonically decreasing
function of $U_0$. It has some similar to the fundamental string
stretched between the D8 and $\bar{D8}$-branes in the unwrapping
near-horizon D4-branes background that the length of fundamental
string is a monotonically decreasing function of $U$ but not $U_0$.
In this background, the D8-branes and $\bar{D8}$-branes join
together at the minimal value $U=U_0$ through wormhole and the
chiral symmetry is broken. It can be shown as the way in section
\ref{ssss}. we can also understand it since there is the attractive
force between the D8 and $\bar{D8}$-branes, thus the D8-branes and
$\bar{D8}$-branes connected form is preferred than the separated
case.

As in the subsection \ref{ssss}, we can study the fundamental string
stretched between the D8 and $\bar{D8}$-branes in the background
(\ref{met}) and (\ref{me}). Assuming the fundamental string extends
in the $(x^0, x^2)$ directions. We are only interested in the string
shape in the $(x^2, U)$ directions, then the induced metric of the
fundamental string is \be ds^2 =
-\left(\frac{U}{R}\right)^{5/2}(dx^0)^2 +
\left(\left(\frac{U}{R}\right)^{5/2}f(U)+ \frac{1}{f(U)}
\left(\frac{R}{U}\right)^{5/2}\left(\frac{dU}{dx^2}\right)^2\right)(dx^2)^2\ee
then substituting into the Nambu-Goto action of string, we can get
\be S_{NG} \sim \int dx^2 \left(\left(\frac{U}{R}\right)^5f(U)+
\frac{U'^2}{f(U)}\right)^{1/2} \ee with the $U'=\frac{dU}{dx^2}$.
With boundary condition $U(0)=U_0^{F_1}$ and $U'(0) = 0$, the
$U(x^2)$ satisfies the first order equation  \be \frac{U^{5/2}
f(U)^{1/2}}{\sqrt{1+\left(\frac{R}{U}\right)^5\frac{U'^2}{f(U)^2}}}
= (U_0^{F_1})^{5/2}f(U_0^{F_1})^{1/2} \label{ll}\ee Here the
$U_0^{F_1}$ is the minimal value of $U$ that corresponds to the
$U_0^F$ of the fundamental string in the background (\ref{metric})
in section \ref{ssss}. We can solve the $x^2(U)$ as follows \be
x^2(U) = (U_0^{F_1} R)^{5/2}f(U_0^{F_1})^{1/2}\int_{U_0^{F_1}}^U
\frac{dU}{U^{5/2} f(U)\sqrt{U^5f(U)-(U_0^{F_1})^5 f(U_0^{F_1})}}
\label{vv}\ee For the $U$ larger than the $U_0^{F_1}$ so that the
upper limit of the integral (\ref{vv}) can be taken to infinity and
can define the $ \frac{L}{2}=x^2(\infty)$. From the integral
(\ref{vv}), if $U_0=U_{KK}$, we get  \be
\frac{L}{2}=x^2(\infty)\approx \frac{R^{5/2}}{4(U_0^{F_1})^{3/2}}
\ee Now we compare the minimal value of $U$ for the fundamental
string and the D8-brane. It is \be
\frac{U_0^{F_1}}{U_0}\approx\frac{5}{4}\ee Hence approximately we
can obtain the $U_0^{F_1}$ is larger than $U_0$ as same as in
(\ref{wsxu}). If the $U_0 \rightarrow \infty$, then the fundamental
string runs and disappears to infinity.

We can also discuss the meson spectrum in this configuration as in
\cite{Ts}. To study the fluctuation modes around the D8-branes, we
can determine the degrees of freedom corresponding to the massless
particles. In the paper \cite{Maria}, the authors give some
discussion about the meson spectrum of this model.

Until now, the all above discussion is at zero temperature. In the
following we turn to study the theory in the strong 't Hooft
coupling region and at a finite temperature. We do a analytically
continue the time coordinate $ x^0 \rightarrow x^0_E = i x^0 $ with
period $T, ~~ x^0_E \approx x^0_E + T $, and impose antiperiodic
boundary conditions on the fermions in the direction $x^0_E$
\cite{Ew}. Then there are two Euclidean supergravity solutions which
need to considered that it is same as the D4-branes \cite{Mk,Rd}.
The first one is \be ds^2= \left(\frac{U}{R}\right)^{5/2}((dx^0_E)^2
+ (dx^1)^2 +
f(U)(dx^2)^2)+\left(\frac{R}{U}\right)^{5/2}(\frac{dU^2}{f(U)}+U^2d\Omega_6^2)
\label{me1} \ee which is obvious the Euclideanised version of metric
(\ref{met}). And from the black D2-branes, the second metric can be
written as \be ds^2=
\left(\frac{U}{R}\right)^{5/2}(\tilde{f}(U)(dx^0_E)^2 +
(dx^1)^2+(dx^2)^2)+\left(\frac{R}{U}\right)^{5/2}(\frac{dU^2}{\tilde{f}(U)}+U^2d\Omega_6^2)
\label{me2} \ee where the  $\tilde{f}(U)= 1-
\left(\frac{U_T}{U}\right)^5$. These two metric are equal to each
other after commuting $x^0_E$ with $x^2$ and $U_{KK}$ with $U_T$. To
avoid the singularity of the metrics at $U= U_{KK}$ and $U_T$, the
coordinates must satisfy the periodic boundary conditions. Then the
periodicity of $\delta x^2$ in (\ref{me1}) satisfies (\ref{antip}),
and in (\ref{me2}) the periodicity of $x^0_E$ is $\delta x^0_E =
\frac{4\pi R^{5/2}}{5U_T^{3/2}}$. However, the periodicity of the
other two coordinates is arbitrary. Hence, the temperature of
background (\ref{me1}) is arbitrary, the second one is proportional
to the inverse of $\delta x^0_E$. At a given temperature, we need to
decide which background dominates the path integral. We can compare
the free energies of these two backgrounds and see which one has a
lower free energy. There is a critical temperature $T^*$, under it,
the background (\ref{me1}) dominates the path integral. Otherwise,
if the temperature is higher than the critical temperature $T^*$,
the metric (\ref{me2}) dominates. From the low temperature to the
high energy temperature, there is a Hawking-Page phase transition
between them \cite{Wh}.

In the low temperature phase, the analysis is same as the zero
temperature case except that the time is Euclidean and compactified
with a circumference $\frac{1}{T}$, where $T$ is the temperature. We
can use the dual description and find the chiral condensate as in
the zero temperature case. The equation of motion of D8-brane in
this background is similar to the equation (\ref{aaa}). We can find
that the configuration of the D8-branes and $\bar{D8}$-branes are
connected each other in this background. It hints the chiral
symmetry breaking in the dual gauge theory.

However, in the high temperature phase, the background (\ref{me2})
dominates the path integral. As done in the zero temperature, we use
the D8-brane to probe the background and find the dynamical curve in
the plane $(x^2, U)$. Here we assume that the D8-brane extends in
directions $(x^0_E,x^1,U, \Omega_6 )$, the induced metric on the
D8-brane is \be \begin{array}{lll} ds^2 &=
\left(\frac{U}{R}\right)^{5/2}(\tilde{f}(U)(dx^0_E)^2 + (dx^1)^2)\\
&+\left(\left(\frac{U}{R}\right)^{5/2}\left(\frac{dx^2}{dU}\right)^2+
\frac{1}{\tilde{f}(U)}\left(\frac{R}{U}\right)^{5/2}\right)dU^2+
\left(\frac{R}{U}\right)^{5/2}U^2d\Omega_6^2 \end{array}
\label{indu3}\ee Then using the dilaton (\ref{me}) and induced
metric (\ref{indu3}) and neglecting some proportional constant, the
DBI action is \be S \sim \int dx^2 U
\sqrt{\left(\frac{U}{R}\right)^5\tilde{f}(U)+ U'^2} \label{nn}\ee
where the $U'= \frac{dU}{dx^2}$. Obviously, this equation will
reduce to equation (\ref{act}) if $\tilde{f}(U)=1$, i.e.  $U_T =0 $.
From the above action (\ref{nn}), we can get the equation of motion
\be
\left(\frac{d}{dx^2}\right)\left(\frac{U^6\tilde{f}(U)}{\sqrt{\left(\frac{U}{R}\right)^5\tilde{f}(U)+U'^2}}\right)
= 0 \ee If we set the boundary condition of the above equation as
$U(x^2)=U_0$ and  $U'(x^2)= 0$, then we obtain the first order
equation \be
\frac{U^{7/2}\tilde{f}(U)^{1/2}}{\sqrt{1+\left(\frac{R}{U}\right)^5\frac{U'^2}{\tilde{f}(U)}}}
=U_0^{7/2}\tilde{f}(U_0)^{1/2} \ee  Then the curve $x^2(U)$ can be
expressed as \be x^2(U) = \int_{U_0}^U
\frac{dU}{\tilde{f}(U)^{1/2}\left(\frac{U}{R}\right)^{5/2}
\sqrt{\left(\frac{U}{U_0}\right)^5\frac{\tilde{f}(U)}{\tilde{f}(U_0)}-1}}\label{a1}\ee
The value $U_0$ at $x^2=0$ is the smallest value of $U$ where the
D8-branes are connected with the $\bar{D8}$-branes.  When the
minimal value $U_0$ of the D8-brane shape $U(x^2)$, $U_0 \rightarrow
U_T$, then the D8-brane approaches to the horizon of the black hole.
If $U_0$ reaches to the value $U_T$, this means that the wormhole of
the connected D8-$\bar{D8}$-brane pairs reaches the horizon of this
black hole background. However, if the $U_0\rightarrow \infty$, then
the D8-$\bar{D8}$-brane pairs approach to infinity and disappear.

\begin{figure}[h]
\centering{\includegraphics[]{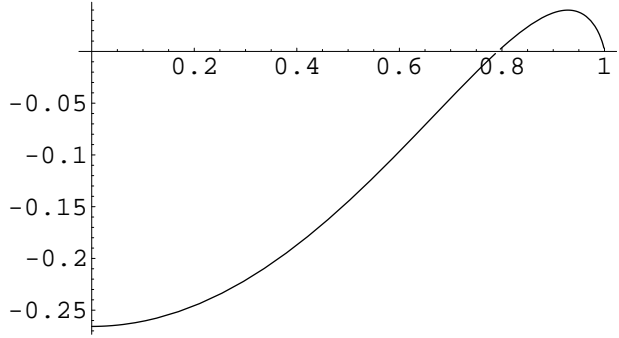}}
 \caption{ the relations between the  energy difference $\delta$S and $ y_T$.}
 \label{gn1}
 \end{figure}

Substituting the equation (\ref{a1}) into the DBI action (\ref{nn})
and using the integral variable $y\equiv\frac{U}{U_0}$, we get the
\be S_1\sim \int_1^\infty dy
\frac{y}{\sqrt{1-\frac{\tilde{f}(1)}{\tilde{f}(y)y^7}}}\label{a2}\ee
For the D8 and $\bar{D8}$-branes separated case, we also can get \be
S_2\sim \int_{y_T}^1 dy y+\int_1^\infty dy y \label{a3}\ee  To
determine whether the chiral symmetry is broken or not in high
temperature. We now need to compare the free energy of this
configuration with the energy of the separated D8-branes and
$\bar{D8}$-branes configuration. Since the D8-branes extend to
infinity, thus the energy is divergent. But the difference between
these two configurations is finite. Then from the equation
(\ref{a2}) and (\ref{a3}), the difference energy is proportional to
\be \delta S = S_1-S_2= \int_1^\infty dy
y\left(\frac{1}{\sqrt{1-\frac{\tilde{f}(1)}{\tilde{f}(y)y^7}}}-1\right)
- \int_{y_T}^1 dy y  \ee  if we define the $ z\equiv y^{-7}$, then
the $\delta S$ can be written as \be \delta S = \frac{1}{7}\int_0^1
dz z^{-9/7}\left(
\frac{\sqrt{1-y_T^5z^{5/7}}}{\sqrt{1-y_T^5z^{5/7}-(1-y_T^5)z}}-1\right)-
\frac{1}{2}(1-y_T^2)\label{a4}\ee Then we can use a numerical way to
analyze the integral and draw a Fig. \ref{gn1} in the following

In the following, we can see, in the high temperature phase $T >
\frac{1}{2\pi R_2}$ which is dominated by the background
(\ref{me2}), there exists a first order phase transition of the
chiral symmetry breaking. From the figure \ref{gn1}, there is a
critical point $y_T\sim 0.80$ where  $\delta S = 0$. By increasing
the $y_T$, if the $y_T$ is larger than a critical value $y_T\sim
0.80$, then the energy difference $\delta S$ is positive. Thus the
preferred branes configuration is the disconnected D8 and
$\bar{D8}$-branes case. It means that the chiral symmetry is
unbroken in this region. Otherwise, in the region $y_T< 0.80$, the
configuration of the connected D8 and $\bar{D8}$-branes is
preferred, in this case the chiral symmetry is broken. Therefore, we
know that at the critical point $y_T\sim 0.80$, there is a first
order phase transition of the chiral symmetry breaking. The results
in here is very similar to the case in SS model at finite
temperature that is analyzed by \cite{Js1,Ap}. In the SS model,
there also exists a critical temperature. And above this
temperature, the chiral symmetry is unbroken. However, under it, the
symmetry is broken.

In the equation (\ref{a1}), if  approximately choosing the $U$
larger than the $U_0$, then the upper limit of the integral can be
leaded to infinity. The asymptotic distance $L$ between D8-branes
and $\bar{D8}$-branes is equal to \be \frac{L}{2}=R^{5/2}
\int_{U_0}^\infty
\frac{dU}{\tilde{f}(U)^{1/2}U^{5/2}\sqrt{\left(\frac{U}{U_0}\right)^7\frac{\tilde{f}(U)}{\tilde{f}(U_0)}-1}}\label{dd1}\ee
After a change of the integral variable, the above equation
(\ref{dd1}) becomes \be\begin{array}{lll} \frac{L}{2}=&
U_0^{-3/2}R^{5/2}\int_1^\infty
\frac{dy}{\tilde{f}(y)^{1/2}y^{5/2}\sqrt{\frac{\tilde{f}(y)}{\tilde{f}(1)}y^7-1}}\label{dd2}\\&
=\frac{1}{7}U_0^{-3/2}R^{5/2}\sqrt{1-y_T^5}\int_0^1\frac{dz}{z^{2/7}(1-y_T^5z^{5/7})^{1/2}\sqrt{1-y_T^5z^{5/7}-(1-y_T^5)z}}\end{array}\label{ddd2}
\ee If the $y_T\ll 1 $, the $U_0$ has the relation $U_0^3\sim
R^5/L^2$. This means that the $L\rightarrow 0$ with increasing the
$U_0$, and the D8-$\bar{D8}$-branes will run to infinity and
disappear. When the $y_T\rightarrow 1$, then the
D8-$\bar{D8}$-branes throat will run to the horizon of the black
hole background (\ref{me2}) and eventually meet each other. If we
choose that the $y_T $ is equal to the critical point of the phase
transition of the chiral symmetry. From the integral (\ref{ddd2}),
we get the correspondent value $L\simeq 0.33
\frac{R^{5/2}}{U_0^{3/2}}$ at this critical temperature. Using the
period condition $\delta x^0_E = \frac{4\pi R^{5/2}}{5U_T^{3/2}}$,
we get the relation about the temperature $T$ with the asymptotic
distance $L$ at this critical point, and that is $T^\chi\simeq
\frac{0.09}{L}$. We know that there is a Hawking-Page transition
from the low temperature (\ref{me1}) to the high temperature
(\ref{me2}). And we can set that the critical temperature of
Hawking-Page phase transition is $T^*= \frac{1}{2\pi R_2}$. If the
$L>0.57 R_2 $, then the temperature in this deconfined phase is
always higher than $T^\chi$, thus the deconfinement and chiral
symmetry restoration happen together. On the opposite side $L < 0.57
R_2$, It means that the theory lies in the deconfinement phase and
the deconfinement and chiral symmetry restoration transitions happen
separated.

We have studied the zero and finite temperature case of this
D2-D8-$\bar{D8}$-branes configuration. In the following we give some
discussions about the relations of this branes system with others
D-branes configuration. After performing a T-dual to this D-branes
configuration along the direction $x^2$, we can get a
D1-D9-$\bar{D9}$-branes configuration in the IIB string theory. This
configuration has been studied in \cite{Su}. In this
D1-D9-$\bar{D9}$-branes configuration the D1, D9 and
$\bar{D9}$-branes are parallel each other. As in the section
\ref{config}, we can analyze the massless degrees of freedom on the
D-branes. If choosing the color index $N_c=1$, the low energy
dynamics on the D1-brane is a two-dimensional massless QED(Schwinger
model) which is exactly solvable. This theory is equivalent to a
free massive scalar field theory by using the bosonization
techniques. For the bulk tachyon modes of the 9-$\bar{9}$-string,
the tachyon fields will couple to the fermions on the D1-brane via
Yukawa coupling and the fermions get mass. Then the dynamics on the
D1-brane is a massive two-dimensional QED, it is no longer a
solvable theory. But we can still use the bosonization way to study
it. When the tachyon fields get large vacuum expectation value, the
Yukawa coupling between the bulk tachyon and the worldvolume
fermions of D1-brane is decoupled from the low energy physics, and
the massive two-dimensional QED will reduce to the massless
two-dimensional QED. Also if compactifying one longitudinal
direction of the D1-brane on a circle and performing a duality again
along this compacted direction, we get the D0-D8-$\bar{D8}$-branes
configuration. The D0-branes in D8-$\bar{D8}$-branes background can
be described by the Matrix-theory \cite{Su}. The other brane system
such as D2-D6-$\bar{D6}$ and D2-D4-$\bar{D4}$ brane configurations,
they also have two intersectional dimensions In these brane
configurations \cite{Maria} as the D2-D8-$\bar{D8}$ brane
configuration. One can use the same way to study these brane
configurations as the D2-D8-$\bar{D8}$ brane configuration in this
paper.

\section{summaries}
Motivated by the ads/cft, much work has been done to give a
holographic dual description of the large $N_c$ QCD with flavors by
string theory such as \cite{Ka,Mk,Ts}. In this paper, we construct a
two-dimensional toy model from the D2-D8-$\bar{D8}$-branes
configuration in IIA string theory. In section \ref{config}, we
analyze the massless spectrum on the two-dimensional intersection of
this branes configuration and show them in the table \ref{sdfdfs}.
In the weak coupling region $l_s\ll L\ll \frac{1}{\lambda}$, the low
energy dynamics on the intersection is a non-local GN model. We know
that the GN model is asymptotically free and exactly solvable at
large $N_c$ and finite $N_f$ and dynamical breaks the chiral flavor
symmetry $U(N_f)_L \times U(N_f)_R$. The non-local GN model is a
non-local generalization of the  GN model, it can be seen from the
equation (\ref{effective4}). Using this NGN model, we find the
quarks condensate and the chiral symmetry breaking. But in the
opposite coupling region (\ref{strong}), the interaction becomes
strong and we can't use the perturbative way to study the low energy
dynamics of this brane system. Instead the IR dynamics can be
described by the SUGRA approximation. We find the connected
D8-$\bar{D8}$-branes configuration by the wormhole is preferred. And
it means that the chiral symmetry is broken. This gives a geometric
realization of the chiral symmetry breaking. In both the
descriptions, we find the chiral symmetry is dynamically broken. We
also discuss some bulk and boundary correspondences of this brane
configuration, however, the details of the correspondences is needed
to further study. For the fundamental string in the background
(\ref{metric}), we find some interesting results. From the equation
(\ref{wsxu}), we can see that the minimal value $U_0^F$ of the
fundamental string is always larger than the correspondence one of
the D8-$\bar{D8}$-branes in this background. It means that the
fundamental strings don't reach the minimal value $U_0$ of D8-branes
and stay a short distance away from it. Also we have compared the
strength between the curved and the straight fundamental string. We
find the curved string is longer than the straight one.

In the section \ref{gera}, we give a generalization of the D-branes
configuration. It can be got through wrapping one direction $x^2$ of
the D2-brane on a circle of radius $R_2$ with anti-periodic
condition of the fermions. Then the fermions and scalars get mass
and decouple from the IR physics. The only massless modes on
D2-branes are the quarks and gauge fields. In the weak coupling
region $l_s\ll L\ll R_2 \ll \frac{1}{\lambda}$, the low energy two
dimensional physics can be described by the $QCD_2$. We can use the
standard techniques to study this theory. However, in the opposite
region $ \frac{1}{\lambda} \ll l_s\ll L\ll R_2 $, the interaction
becomes strong. And instead the perturbative analysis, the SUGRA can
be used to describe this region and we can find the chiral symmetry
is dynamically broken as in the section \ref{config}. Following
\cite{Ts}, the meson spectrum of this model need to work out. In
\cite{Maria}, there are some discussions about the meson spectrum of
this model.

In the following we also study the properties of this system at
finite temperature. At low temperature, the metric (\ref{me1})
determines the path integral. However, at high temperature, the
background (\ref{me2}) determines. We know that with the
increasing temperature there happens a Hawking-Page deconfinement
phase transition from the background (\ref{me1}) to (\ref{me2}).
At the high temperature phase, there is a chiral symmetry phase
transition critical point $y_T = 0.80$ or $T^\chi =
\frac{0.09}{L}$. Below this point, the chiral symmetry is broken.
However, up this point, the chiral symmetry is restored. In the
last, we have discussed some relations about this configuration
with the other branes configuration. For example the
D1-D9-$\bar{D9}$-branes configuration, the low energy dynamics on
one D1-brane is two-dimensional QED. If performing the T-duality
again along the D1-branes worldvolume, the brane configuration
becomes the D0-D8-$\bar{D8}$ brane system and the D0-branes
dynamics have a matrix description. There are some other D-brane
configurations such as D2-D6-$\bar{D6}$ and D2-D4-$\bar{D4}$ which
also exist two dimensional intersection. For these brane system,
we can perform the same analysis as the D2-D8-$\bar{D8}$ brane
configuration in this paper.
\\
\\
{\bf Acknowledgements} \\  W. Xu would like to thank K. Ke, M.X. Luo
and H.X. Yang for reading the manuscript and giving many
suggestions,  and L.M. Cao, D.W. Pang, J.H. She, J.P. Shock, X. Su and Z. Wei
for some useful helps and nice discussions.


\end{document}